\documentclass[aps, prl, reprint, superscriptaddress,  floatfix]{revtex4-2}
\usepackage{latexsym}
\usepackage{amsthm}
\usepackage{amssymb}
\usepackage{amsmath}
\usepackage[all]{xy}
\usepackage{float}
\usepackage{graphicx}
\usepackage{subfigure}
\usepackage{dcolumn}
\usepackage{bm}
\usepackage{bbm}
\usepackage[usenames, dvipsnames]{xcolor}
\usepackage{amsfonts}
\usepackage{cases}
\usepackage{multirow}
\usepackage{bigstrut}

\usepackage{xcolor}
\usepackage[normalem]{ulem}

\usepackage{booktabs} 
\usepackage{siunitx}  
\usepackage[pagebackref=false, colorlinks=true,
linkcolor=blue, citecolor=blue,urlcolor=blue,]{hyperref}

\begin{document}

\title{ Stringent  Constraints on Spin-Spin-Velocity-Dependent Exotic Interactions  with a Levitated Magnet Force Sensor}

\author{Kenan Tian}
\affiliation{National Laboratory of Solid State Microstructures and Department of Physics,  Nanjing University, Nanjing 210093,
	China}

\author{Siwen Chen}
\affiliation{National Laboratory of Solid State Microstructures and Department of Physics, Nanjing University, Nanjing 210093,
	China}

\author{Lei Wang}
\affiliation{National Laboratory of Solid State Microstructures and Department of Physics, Nanjing University, Nanjing 210093,
	China}

\author{Yuanji Sheng}
\affiliation{National Laboratory of Solid State Microstructures and Department of Physics, Nanjing University, Nanjing 210093,
	China}

\author{Dingjiang Long}
\affiliation{National Laboratory of Solid State Microstructures and Department of Physics, Nanjing University, Nanjing 210093,
	China}

\author{Rui Li}
\affiliation{National Laboratory of Solid State Microstructures and Department of Physics, Nanjing University, Nanjing 210093,
	China} 

\author{Han Xie}
\affiliation{National Laboratory of Solid State Microstructures and Department of Physics, Nanjing University, Nanjing 210093,
	China}

 \author{Yiming Chen}
 \affiliation{National Laboratory of Solid State Microstructures and Department of Physics,  Nanjing University, Nanjing 210093,
	China}

 \author{Xiang Bian}
\affiliation{National Laboratory of Solid State Microstructures and Department of Physics, Nanjing University, Nanjing 210093,
	China}

 \author{Hao Wang}
\affiliation{National Laboratory of Solid State Microstructures and Department of Physics, Nanjing University, Nanjing 210093,
	China}

\author{Ruoyu Ding}
\affiliation{National Laboratory of Solid State Microstructures and Department of Physics, Nanjing University, Nanjing 210093,
	China}

 \author{Chang-Kui Duan}
  \affiliation{Laboratory of Spin Magnetic Resonance, School of Physical Sciences, Anhui Province Key Laboratory of Scientific Instrument Development and Application, University of Science and Technology of China, Hefei 230026, China}
\affiliation{Hefei National Laboratory, University of Science and Technology of China, Hefei 230088, China }

\author{Peiran Yin}
\affiliation{National Laboratory of Solid State Microstructures and Department of Physics, Nanjing University, Nanjing 210093,
	China}

 \author{Xi Kong}
  \email{kongxi@nju.edu.cn}
\affiliation{National Laboratory of Solid State Microstructures and Department of Physics, Nanjing University, Nanjing 210093,
	China}
 
 \author{Pu Huang}
 \email{hp@nju.edu.cn}
\affiliation{National Laboratory of Solid State Microstructures and Department of Physics, Nanjing University, Nanjing 210093,
	China}

\begin{abstract}
 Exotic spin-spin-velocity-dependent interactions, predicted in extensions of the Standard Model involving new bosonic fields, could resolve fundamental puzzles from dark matter to cosmic asymmetry. However, exploring these weak potential interactions at centimeter scales presents formidable challenges, primarily due to the overwhelming dominance of electromagnetic backgrounds that can easily obscure the weak exotic signals. Here, we utilize a levitated magnet force sensor with ultrahigh electron spin density to probe these interactions. We constrain two interactions individually through a designed spin source and a multi-layer magnetic shielding system that suppresses electromagnetic backgrounds. In this study, we constrain two types of interactions: the \( V_6 \) potential at force ranges from \( 10^{-3} \)\,m to   $ 6 \times 10^{-2} $ m and the \( V_{14} \) potential at ranges greater than \( 10^{-3} \) m. Our measurements establish 95\% confidence-level bounds of \( |f_6| \leq 2.12 \times 10^{-13} \) and \( |f_{14}| \leq 2.34 \times 10^{-23} \) at $\lambda = 1.6 \times 10^{-2}$ m, improving prior limits by up to 12 and 13 orders of magnitude, respectively. Our result demonstrates the levitated magnet as a highly sensitive probe for detecting new bosonic fields in extensions of the Standard Model.

\end{abstract}

\maketitle
\textit{Introduction}\textbf{---}As the cornerstone of particle physics, the Standard Model successfully unifies the electromagnetic, weak, and strong interactions while accurately predicting fundamental particles—including the Higgs boson—to establish a robust theoretical framework for matter's microscopic structure. However, it fails to explain cosmological phenomena including dark matter and the observed matter-antimatter asymmetry, motivating searches for new particles and interactions beyond the Standard Model \cite{cong2024spindependentexoticinteractions}. Exemplary predictions include Moody and Wilczek's 16 types of axion-mediated spin-dependent exotic interactions between fermions \cite{PhysRevD.30.130}.  Searching for exotic spin dependent interactions, particularly at the low-energy frontier, in precise experiments has attracted broad interest in recent years \cite{PhysRevD.78.092006,PhysRevLett.108.181801,science.1227460,PhysRevLett.111.151802,PhysRevLett.115.081801,PhysRevLett.115.201801,PhysRevD.95.075014,PhysRevLett.121.080402,NatCommun.10.2245,PhysRevLett.124.161801,PhysRevLett.125.201802,sciadv.abi9535,PhysRevLett.134.111001,PhysRevLett.127.010501,NationalScienceReview.10.nwac262,PhysRevLett.129.051801,PhysRevLett.130.133202,PhysRevLett.121.261803,PhysRevLett.132.180801}.

These potentials arise from linear combinations of six fundamental coupling constants: \(g_{\mathrm{S}}\), \(g_{\mathrm{P}}\), \(g_{\mathrm{V}}\), \(g_{\mathrm{A}}\), \(\text{Re}(C)\), and \(\text{Im}(C)\), which parametrize the strength of scalar, pseudoscalar, vector, axial-vector, tensor, and pseudotensor interactions, respectively \cite{PhysRevA.99.022113}. These couplings define how Standard Model fermions interact with new bosonic fields, such as spin-0 bosons (via \(g_{\mathrm{S}}^e, g_{\mathrm{P}}^e\)) or spin-1 bosons (via \(g_{\mathrm{V}}^e, g_{\mathrm{A}}^e, \operatorname{Re}(C_e), \operatorname{Im}(C_e)\))) \cite{JournalofHighEnergyPhysics.2006.005}.

The macroscopic potentials \(V_6\) and \(V_{14}\) investigated in our experiment are manifestations of specific combinations of these microscopic couplings \cite{PhysRevA.99.022113}. Constraining the parameters \(f_6\) and \(f_{14}\) thus directly translates into limits on the underlying electron-boson coupling constants, such as the axial-vector and pseudotensor couplings for \(V_6\) (\(g_{\mathrm{A}}^{e}\) and \(\text{Im}(\mathcal{C}_e)\)), and the scalar and pseudoscalar couplings for \(V_{14}\) (\(g_{\mathrm{S}}^{e}\) and $g_{\mathrm{P}}^{e}$).

Levitation-based force sensors—including diamagnetic \cite{NewJPhys.20.063028,PhysRevApplied.15.024061,ApplPhysLett.12.124002,NatPhys.18.1181–1185,ji2025levitatedsensormagnetometryambient}, superconducting \cite{ApplPhysLett.115.224101,PhysRevLett.131.043603,Sci.Adv.eadk2949}, electric \cite{PhysRevLett.114.123602,PhysRevLett.132.133602}, and optical levitation \cite{NatPhy.9.1745-2481, NatNanotechnol.9.425,science.aba3993,PhysRevLett.128.111101} systems—provide a promising platform due to their ultrahigh force sensitivity. These sensors hold great potential in a variety of emerging applications such as dark matter detection \cite{PhysRevLett.125.181102,PhysRevLett.134.181402}, gravitational waves \cite{PhysRevLett.134.181402} and wave function collapse \cite{PhysRevResearch.2.023349}. Diamagnetic levitation has been successfully applied in explorations of dark matter and dark energy. Building on this platform, a previous study \cite{PhysRevLett.134.111001} reported an experimental setup utilizing the levitation of non-magnetic materials to investigate the spin-velocity-dependent interaction.

Among the various beyond-Standard-Model proposals, exotic spin-spin-velocity-dependent interactions—mediated by hypothetical particles such as $Z'$ bosons or paraphotons \cite{PhysRevA.99.022113}—are particularly compelling candidates to address these outstanding issues. Direct experimental searches for the forces arising from these exchanges are therefore crucial. Atomic magnetometers \cite{PhysRevLett.121.261803} and NV centers \cite{PhysRevLett.132.180801} have constrained the relevant parameter space for exotic spin-spin-velocity-dependent interactions. It is noteworthy that the detection sensitivity for such exotic interactions can be significantly improved when both sensor and source consist of high-density spins. However, in those systems,  where spins serve as the magnetic sensors, the achievable spin density within their effective detection volumes remains relatively limited.

In this Letter, we utilize a levitated magnet with ultrahigh electron spin density as a force sensor to probe exotic spin-spin-velocity-dependent interactions. With a magnetic shielding system that suppresses conventional electromagnetic backgrounds, we probe two types of spin-spin-velocity-dependent interactions at the centimeter scale. The resulting constraints on the coupling coefficients $|f_6|$ and $|f_{14}|$ for these exotic interactions at $\lambda = 1.6 \times 10^{-2}$ m significantly surpass previous results by over 12 and 13 orders of magnitude, respectively.

\begin{figure*}[htbp]
    \centering
    \hypertarget{exp-sys1}{}
    \includegraphics[width=\textwidth,height=0.32\textheight]{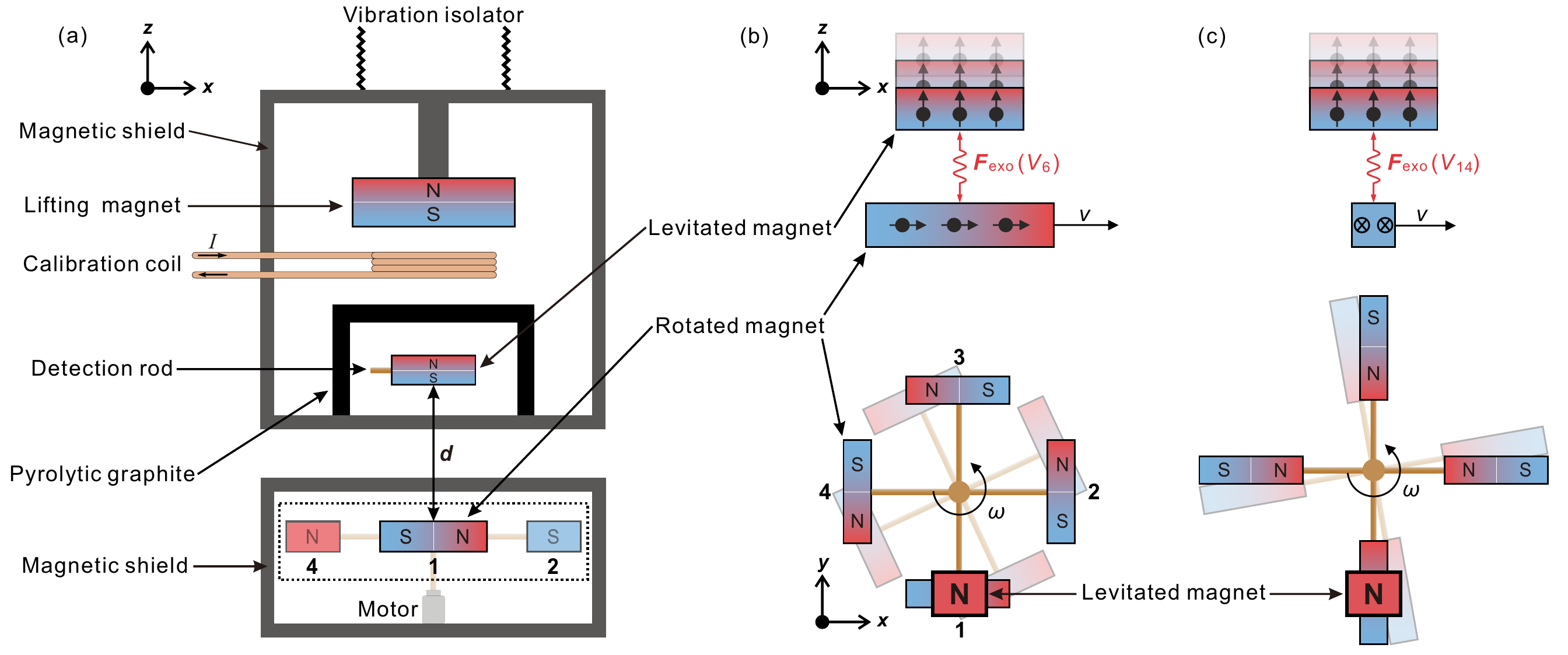}
    \caption{(Color online). Schematic of experimental setup. 
   (a) The force sensor consists of a levitated magnet and a detection rod. The detection rod monitors the displacement of the force sensor. Inside the black dashed box is a magnet turntable used in the $V_6$ experiment, which comprises four magnets symmetrically arranged around the rotation axis and is driven by a motor. When the magnet turntable rotates, the exotic interaction force arises between the levitated magnet (the force sensor) and the rotated magnets, exerting a periodic exotic interaction on the force sensor. The lifting  magnet and the pyrolytic graphite together provide a stable condition for the levitated magnet (the force sensor). The excitation coil is used to apply a magnetic signal of known strength to calibrate the experimental signals. The magnetic shields are employed to suppress electromagnetic interactions between the levitated magnet (the force sensor) and the rotated magnets. (b) The upper half of the figure shows the electron spin orientations within both magnets for $V_{6}$ experiment, and the lower half provides a top view of the magnet turntable for $V_{6}$ experiment. (c) Experimental setup for $V_{14}$ experiment.}
   \label{fig:exp-sys}
\end{figure*}

\textit{Experimental principle and system}\textbf{---}We performed two individual experiments to probe distinct types of exotic interactions described by the potential:
\begin{align}
\label{displacement1}
V_6 =\; & -f_6 \frac{\hbar^2}{4\pi m_e c} \Bigl[
(\hat{\boldsymbol{\sigma}}_1 \cdot \boldsymbol{v})(\hat{\boldsymbol{\sigma}}_2 \cdot \hat{\boldsymbol{r}}) \nonumber \\
& + (\hat{\boldsymbol{\sigma}}_1 \cdot \hat{\boldsymbol{r}})(\hat{\boldsymbol{\sigma}}_2 \cdot \boldsymbol{v})
\Bigr]
\left( \frac{1}{\lambda r} + \frac{1}{r^2} \right) e^{-r/\lambda},
\end{align}
\begin{equation}
\label{displacement2}
V_{14} = f_{14} \frac{\hbar}{4\pi}
\left[ (\hat{\boldsymbol{\sigma}}_1 \times \hat{\boldsymbol{\sigma}}_2) \cdot \boldsymbol{v} \right]
\left( \frac{1}{r} \right) e^{-r/\lambda}.
\end{equation}
Here, $f_6$ originates from interactions involving the axial-vector $g^e_\text{A}$ and pseudotensor $(\text{Im}\mathcal{C}_e)$, and $f_{14}$ arises from mechanisms dependent on the scalar $g^e_\text{S}$ and pseudoscalar $g^e_\text{P}$ couplings. $\hbar$ is the reduced Planck constant, and $\hat{\boldsymbol{r}} = \boldsymbol{r}/|\boldsymbol{r}|$ is a unit vector representing the relative position between two spins. The constants $c$ and $m_e$ represent the speed of light in vacuum and the electron mass, respectively. The interaction range is defined as $\lambda = \hbar / (m_b c)$, where $m_b$ is the mass of the mediating boson. The spin directions of the two sources are denoted by $\hat{\boldsymbol{\sigma}}_1$ and $\hat{\boldsymbol{\sigma}}_2$, and their relative velocity is described by $\boldsymbol{v}$.

Figure \hyperlink{exp-sys1}{1(a)} shows the schematic of the experimental setup. 
The force detection system is briefly outlined in the upper  of Fig.\,\hyperlink{exp-sys1}{1(a)}  with its key component being a levitated vibration sensor that exhibits high sensitivity to forces \cite{PhysRevLett.134.111001}. The force sensor consists of a levitated magnet and a detection rod. Situated above the levitated magnet is another magnet separated by pyrolytic graphite. The stable suspension of the levitated magnet relies on a delicate equilibrium among three forces: the upward magnetic force from the lifting magnet, the downward diamagnetic force induced by the pyrolytic graphite, and gravity. The detection rod is coupled to the levitated magnet to track its motion by detecting variations in light intensity when the rod obstructs the light beam partially. Consequently, vibration signals can be captured through the  detection rod. This configuration exhibits a remarkable sensitivity to resonant disturbances. To mitigate vibrations originating from the ground, the whole system is set up on a vibration isolation platform.

The primary aim of our experiment is to search for the exotic interaction between spins. In addition to the levitated magnet acting as the primary spin source, a rotated magnet is positioned below to serve as the secondary spin source (as depicted in the lower segment of Fig.\,\hyperlink{exp-sys1}{1(a)}). In the presence of an exotic interaction, the motion of the levitated magnet is perturbed by this rotated magnet. By modulating this perturbation, the force sensor can accurately quantify it, facilitating precise measurement of the exotic force. The mutual magnetic interaction between the two magnets has been effectively suppressed to a negligible level within the magnetic shield, an observation consistently noted in subsequent experiments.

In order to implement cyclic modulation of exotic interaction forces, the rotated spin source is propelled by a motorized turntable. Figure \hyperlink{exp-sys1}{1(b)} below presents an overhead view for measuring $V_6$ with this setup (as delineated by dashed lines in Fig.\,\hyperlink{exp-sys1}{1(a)}), illustrating a symmetric arrangement of magnets along the rotational axis $z$, with spins aligned perpendicular to the radial direction.
As the turntable rotates, these magnets pass sequentially beneath the force sensor. This motion creates a periodic modulation of the exotic interaction force. When rotated magnet 1 is precisely aligned directly below the sensor (at the 0° position), the force peak is maximized. A 45° rotation from this alignment shifts the gap between rotated magnets 1 and 2 to a position, minimizing the force. Upon reaching a 90° rotation, rotated magnet 2 approaches the sensor closest, leading to a second force peak (the comprehensive quantitative analysis of this periodic force is provided in the Supplementary Material \cite{supplemental}).

During one complete rotation cycle, there are four distinct peaks in force. At an angular velocity of $\omega$, the magnet turntable induces exotic interaction forces with a frequency of 4$\omega$/(2$\pi$) experienced by the force sensor. By precisely regulating the rotational speed of the drive motor, we can continuously tune the frequency of these exotic interaction forces. Resonance occurs when the frequency of these exotic forces aligns with the natural resonant frequency of the force sensor along the $z$-axis, facilitating precise measurement of the exotic interactions. The approach for measuring $V_{14}$ resembles that of $V_{6},$ where the optimal strength of exotic interaction forces is observed when the spins of the two sources are perpendicular, in line with  Eq.~(\ref{displacement2}). Thus, as depicted in Fig. \hyperlink{exp-sys1}{1(c)}, in contrast to the prior scenario, the electron spins in the rotated magnets are oriented radially.

\begin{figure*}[htbp]
    \centering
    \hypertarget{exp-sys2}{}
    \includegraphics[width=\textwidth,height=0.62\textheight]{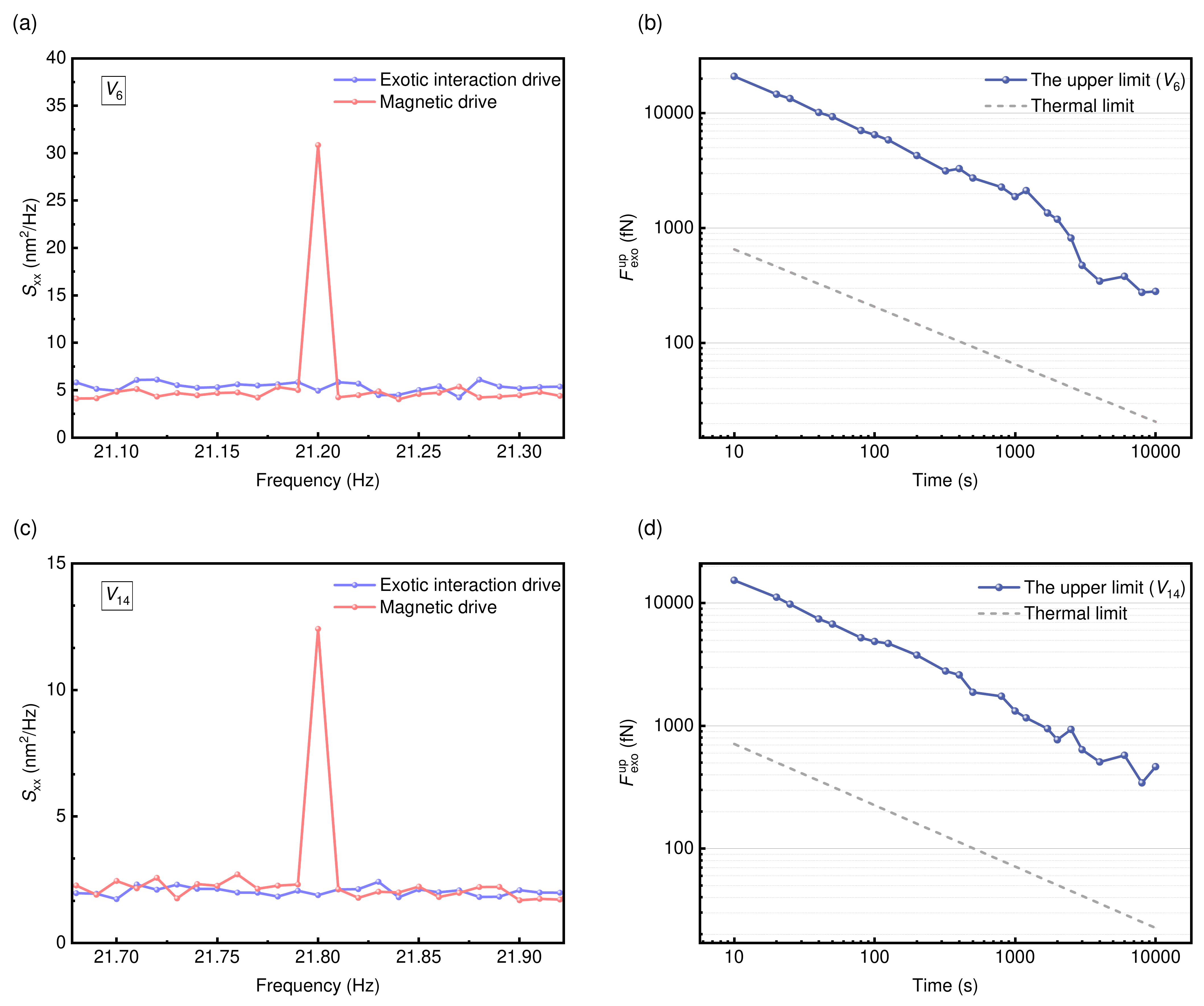}
    \caption{Experimental results. (a) Power spectral density $S_{XX}$ of the levitated magnet displacement in the $V_6$ interaction search. The blue curve represents measurements under excitation by the four-magnet spin source (exotic interaction force), while the red curve shows the reference spectrum driven by a calibrated coil, corresponding to a force strength of $9.00\times10^{-12}\,\text{N}$. We calculate the power spectral density 100 times over a period of 100 s to obtain its average. (b) The blue line shows the upper limit of the exotic interaction with the potential of $V_6$ at the 95\% confidence level, as a function of measurement time. The gray dashed line indicates the theoretical limit set by Brownian thermal noise. (c,d) Corresponding data for the $V_{14}$ interaction.  The calibrated coil driving force is $6.98\times10^{-12}~\text{N}$.}
\end{figure*}

\begin{figure*}[htbp]
    \centering
    \hypertarget{exp-sys3}{}
    \includegraphics[width=\textwidth,]{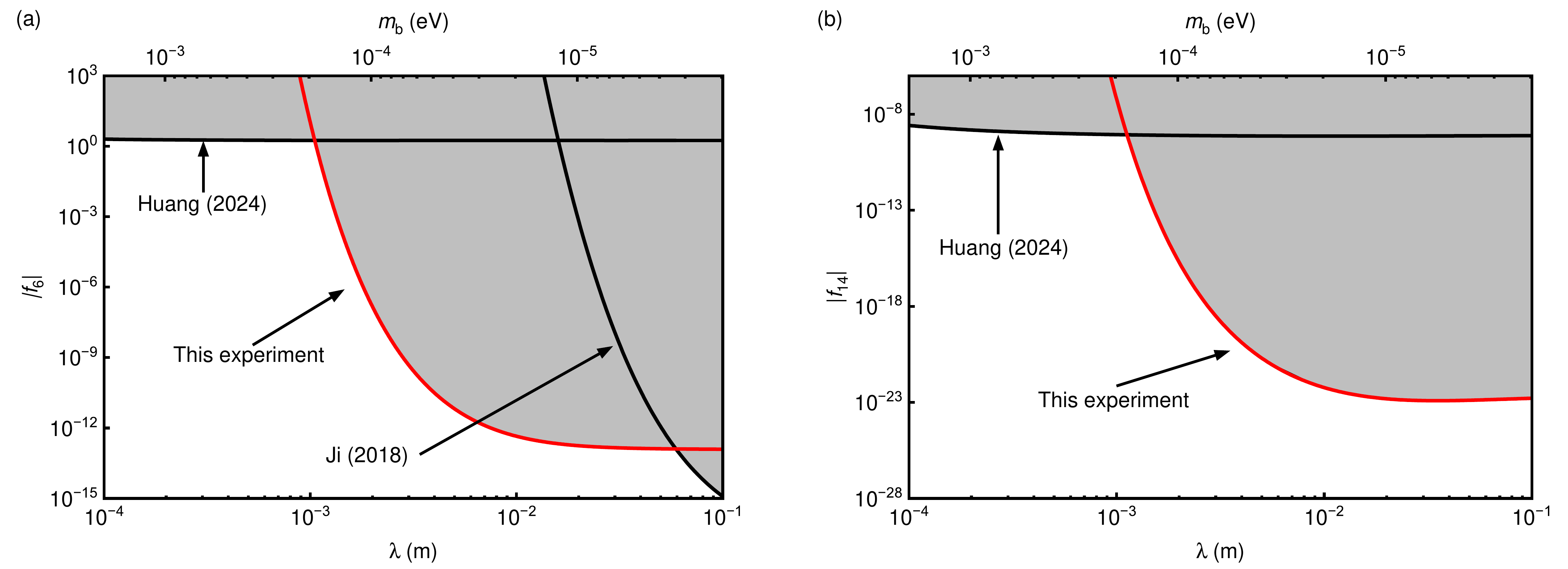}
    \caption{Constraints on the exotic spin-spin-velocity-dependent interactions as a function of the force range $\lambda$ and the mass of the bosons $m_b$. The black lines represent the upper limits established by previous experiments, while the red lines show the upper bounds obtained from our experiment. (a) For $|f_{6}|$, an improved laboratory bound is established in the force range from \( 10^{-3} \)\,m to   $ 6 \times 10^{-2} $ m. At $\lambda = 1.6 \times 10^{-2}$ m, the upper limit of the coupling constant is $|f_6| \leq 2.12 \times 10^{-13}$, which improves the previous result by up to 12 orders of magnitude.  (b) For $|f_{14}|$, the constrained force range extends from $1$ mm to longer distances. At $\lambda = 1.6 \times 10^{-2}$ m, the upper limit of the coupling constant is $|f_{14}| \leq 2.34 \times 10^{-23}$, representing an improvement of 13 orders of magnitude.
}
\end{figure*}

The maximum exotic interaction force between the rotated magnets and the force sensor can be defined as: 
\begin{equation}
\label{displacement5}
F_{\mathrm{exo},z} = - \int_{\rm{RM}}\int_{\rm{Sensor}} \rho_i \rho_j
\frac{\partial V(\mathbf{r}_i, \mathbf{r}_j)}{\partial z_i}\,
d\mathbf{r}_i \, d\mathbf{r}_j,
\end{equation}
 where $-\partial V / \partial z_{i}$ represents the $z$-component of the potential energy gradient $V$, as indicated by Eq.~(\ref{displacement1}) and Eq.~(\ref{displacement2}). The double integrals are performed over the volumes of the rotated magnets (RM) and the force sensor, respectively. Here, $\rho_i$ and $\rho_j$ denote the electron spin densities of the force sensor and the rotated magnets, respectively. The relationship between spin density $\rho$ and magnetization $M$ is given by $\rho = M/\mu_B$, where $\mu_B$ denotes the Bohr magneton.

\textit{Data analysis}\textbf{---}In this experiment, the force sensor is subject not only to the exotic interaction but also to background forces including magnetic and electrostatic forces, external vibrations, and Brownian thermal noise. The sensor's motion is governed by:
\begin{equation}
m\ddot{X}(t) + m\gamma \dot{X}(t) + m\omega_0^2 X(t) = F_{\mathrm{dri}}(t),
\end{equation}
where $m$ is the force sensor's mass, $\gamma$ is the damping rate,  $\omega_0$ is its resonant angular frequency, and $F_{\mathrm{dri}}(t)$ consists of the exotic interaction and all aforementioned background forces.

Here we evaluate the influence of background forces and dissipation on the measurements. A vibration-isolation platform is utilized to minimize external mechanical vibrations, while magnetic shields are applied to reduce the impact of electromagnetic forces. However, the proximity of the vibrating levitated magnet to the magnetic shield causes an increase in eddy current dissipation within the shields. Efforts are made to position the magnetic shields far from the force sensor; however, due to geometric configuration of the setup, further reduction of eddy current dissipation is unfeasible. To evaluate the various noise effects, we characterize noise of the system during the $V_6$ measurement (Table~\hyperref[tab:noise]{I}). Calibration of the intensity-displacement response is carried out using the excitation coil, establishing a quantitative relationship between changes in light intensity and force sensor displacement. Our findings reveal that measurement noise significantly outweighs other noise sources (Table~\hyperref[tab:noise]{I}). Additionally, subsequent experiments demonstrated no  noise effects found at the resonance frequency, indicating negligible influence of background noises like electromagnetic interference on our measurements.

\begin{table}[htbp]
\centering
\caption{Experimental Parameters and Values for $V_6$ Constraints. }
\label{tab:noise}
\begin{tabular}{lccc}
\hline\hline\addlinespace
\textbf{Parameter} & \textbf{$\rm{Value}$} \\[1ex]
\hline\addlinespace

Mechanical dissipation $\gamma_{6}/2\pi$& $0.42$  $\text{Hz}$ \\[1ex]
Resonant frequency $f_{0}$ & $21.20$  $\text{Hz}$ \\[1ex]

Spin density of force sensor $\rho_{\rm{Sensor}}$ & $1.26\times 10^{29}$ $\rm{m^{-3}}$\\[1ex]

Spin density of rotated magnets $\rho_{\rm{RM}}$ & $1.02\times 10^{29}$  $\rm{m^{-3}}$ \\[1ex]

Sensor-rotated magnet distance $d$ & $36.77$ mm \\[1ex]

Thermal noise $\sqrt{S_{\text{th}}}$ & $0.10$  nm/$\sqrt{\text{Hz}}$  \\[1ex]
Measurement noise $\sqrt{S_{\text{mea}}}$ & $2.25$  nm/$\sqrt{\text{Hz}}$  \\[1ex]

\hline\hline
\end{tabular}
\end{table}

Based on the configuration presented in  Figs.\,\hyperlink{exp-sys1}{1(b)} and \hyperlink{exp-sys1}{1(c)}, we conducted two individual experiments to investigate the two exotic interactions denoted as $V_{6}$ and $V_{14}$. Calibration of the measurement absolute values was performed using the reference signals generated by the excitation coil. For the $V_6$ measurement, we first adjusted the magnet turntable speed to match the interaction frequency to the resonant frequency of the sensor's \(z\)-axis translational mode. The sensor displacement under this resonant condition is denoted as $X_1(t)$. In the second measurement, the motor was turned off, and a calibrated signal was applied to the excitation coil at the \(z\)-axis translational mode resonance frequency of the force sensor. The resulting displacement is denoted as $X_2(t)$. The same operational procedures were used for $V_{14}$, except for a modification in the geometrical configuration.

Figure \hyperlink{exp-sys2}{2(a)} illustrates the power spectral density of the displacement signal $X_i(t)$ of the $V_{6}$ experiment, defined as $S^{(i)}_{XX}(\omega) = 2T \langle |X_i^2(\omega)| \rangle$. Upon rotation of the turntable, we conducted measurements of the power spectrum $S^{(1)}_{XX}(\omega)$, encompassing potential exotic interaction signals  (blue curve in Fig.\,\hyperlink{exp-sys2}{2(a)}). Then the rotation of the turntable was halted for the reference signal, with solely the coil excitation signal being utilized to quantify the displacement spectrum  $S^{(2)}_{XX}(\omega)$ (red curve in Fig. \hyperlink{exp-sys2}{2(a)}).

The blue line in Fig.\,\hyperlink{exp-sys2}{2(b)} shows the upper limit of the exotic force $F^{\rm up}_{\rm limit}$ at the 95\% confidence level as a function of the measurement time derived from the blue curve (i.e., the measurement noise) in Fig.\,\hyperlink{exp-sys2}{2(a)}. The gray dashed line shows the thermal noise limit. As the measurement noise is higher than the thermal noise, its associated force limit is higher. Additionally, the full-frequency logarithmic power spectral density  and the details of the system's vibration modes (Fig.\,\hyperlink{exp-sys4}{4} in the End Matter) verify that the $z$-axis translational mode is strictly separated from the other five vibration modes. The upper limit from the measurement noise is found to be \( 2.81 \times 10^{-13}~\mathrm{N} \) for \textit{\( V_6 \)} experiment.  Figures \hyperlink{exp-sys2}{2(c)} and~\hyperlink{exp-sys2}{2(d)} present corresponding data from the \textit{\( V_{14} \)} experiment, and the upper limit is found to be \( 4.65 \times 10^{-13}~\mathrm{N} \) for \textit{\( V_{14} \)} experiment.

Figures \hyperlink{exp-sys3}{3(a)} and~\hyperlink{exp-sys3}{3(b)} present the upper bounds on the spin-spin-velocity-dependent interactions,  $|f_{6}|$ and $|f_{14}|$, respectively, established by this work, compared with constraints from previous experiments. The gray shaded areas represent the parameter space excluded experimentally. The relative systematic errors are 32.20\% for the $V_6$ experiment and 28.97\% for the $V_{14}$ experiment, arising from various experimental factors (see Supplementary Material \cite{supplemental}).

For the interaction coupling $|f_6|$,  at short ranges $\lambda < 10^{-3}$  $\mathrm{m}$, Huang et al. \cite{PhysRevLett.132.180801} set limits using nitrogen-vacancy centers, while at larger ranges $\lambda > 6 \times 10^{-2}$\,m Ji et al. \cite{PhysRevLett.121.261803} probed with atomic magnetometers. Our work covers the gap range $10^{-3}$  $\mathrm{m}$ $ < \lambda < 6 \times 10^{-2} $ $\mathrm{m}$, establishing the most stringent constraints in this range. At $\lambda = 1.6 \times 10^{-2} $ $\mathrm{m}$, our bound on $|f_6|$ improves previous results by over 12 orders of magnitude. For $|f_{14}|$, limits for $\lambda < 10^{-3} $ m were set by Huang et al. \cite{PhysRevLett.132.180801} using nitrogen-vacancy centers. Our experiment extends the constraints to $\lambda > 10^{-3} $ $\mathrm{m}$, providing the strongest bounds in this region. Notably, at $\lambda = 1.6\times 10^{-2}$  $\mathrm{m}$, we improve the limit on $|f_{14}|$ by more than 13 orders of magnitude.

To more directly connect to the fundamental Lagrangian of new physics and to compare our results with constraints from particle physics, we present constraint plots with the fundamental coupling constants  $|g_{\rm A}^{\rm e} \mathrm{Im}(C_e) / \Lambda^{2}|$ and $|g_{\rm P}^{\rm e} g_{\rm S}^{\rm e}|$ on the vertical axis and the interaction range $\lambda$ (or the boson mass $m_\text{b}$) on the horizontal axis, as shown in Fig. \hyperlink{exp-sys5}{5} in the End Matter, where $\Lambda$ is the ultraviolet cutoff scale. At $\lambda = 1.6 \times 10^{-2}$ m, the upper constraint of the coupling constant is $|g_{\rm A}^{\rm e} \mathrm{Im}(C_e) / \Lambda^{2}| \leq 3.17 \times 10^{-31}~\rm{eV}^{-2}$, which improves upon previous constraints by up to 12 orders of magnitude.

\textit{Conclusion}\textbf{---}We report a novel method utilizing a levitated magnet as the force sensor to detect spin-spin-velocity-dependent interactions. Capitalizing on the intrinsic high-density electron spins of the levitated magnet, we improve the resulting constraints on $|f_6|$ and $|f_{14}|$ at $\lambda = 1.6 \times 10^{-2}$ m, surpassing previous results by over 12 and 13 orders of magnitude, respectively. This study provides an innovative experimental approach to probe new physics beyond the Standard Model and opens new avenues for detecting exotic interactions at the centimeter scale.

\textit{Acknowledgments}\textbf{---}This work was supported by the National Natural Science Foundation of China (Grants No.\,T2388102, No.\,12574531, No.\,125704282, No.\,12574531), the Primary Research and Development Plan of Jiangsu Province (Grant No.\,BE2021004-2), the Fundamental Research Funds for the Central Universities (Grant No.\,14380236), and the Nanjing University PhD Student Zhujian Program. We acknowledge support from Jiangsu Key Laboratory of Quantum Information Science and Technology, Nanjing University, China.

K. T., S. C., L. W., and Y. S. contributed equally to this work.

\textit{Data availability}\textbf{---}The data that support the findings of this study are openly available \cite{tian2026raw}.

\bibliographystyle{apsrev4-2}
\bibliography{References}

\appendix
\onecolumngrid                     
\section{End Matter}
\twocolumngrid  

\begin{figure*}[htbp]               
    \centering
     \hypertarget{exp-sys4}{}
    \includegraphics[width=\textwidth,]{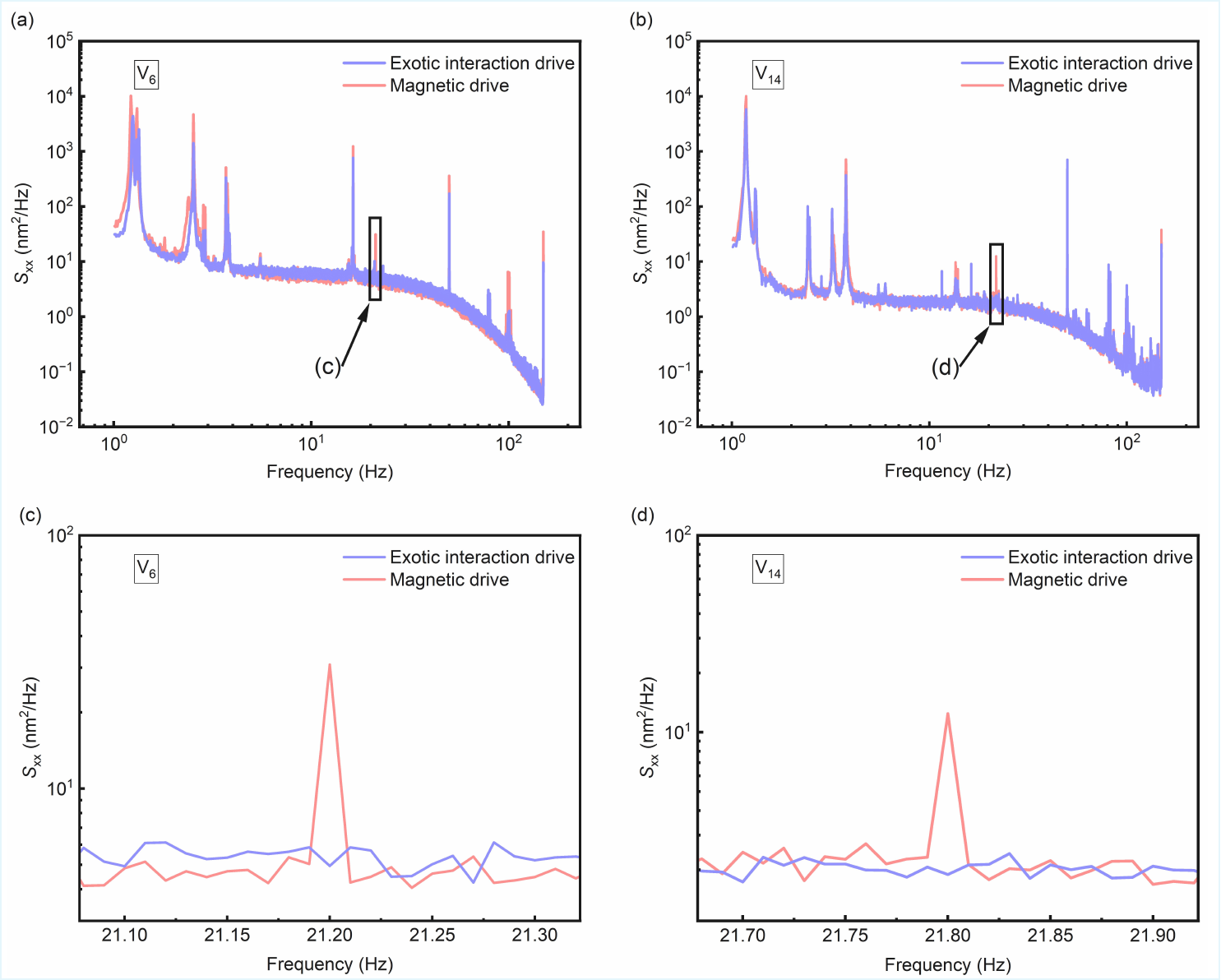}
    \caption{(Color online). Power spectral density $S_{XX}$ of the force sensor displacement in the (a) $V_6$ and (b) $V_{14}$ interaction searches. (c, d) Magnified views of the black-boxed regions in (a) and (b), respectively, showing the frequency range around the driving frequency.}
   
\end{figure*}

\textit{Full-frequency logarithmic power spectral density}\textbf{---}
The power spectral density of the force sensor from 1 to 150 Hz is presented in Fig.~\hyperlink{exp-sys4}{4} on a logarithmic vertical axis, to examine whether a possible signal at the specific frequency might be revealed.
The curve near the resonant frequency is magnified in Fig.~\hyperlink{exp-sys4}{4}(c,d) (same as Fig. 2 in the main text but with a logarithmic scale), confirming the absence of any exotic peaks at the target frequency above the noise floor.

\begin{table}[!ht]
\centering
\caption{Frequencies and linewidths of the six degrees of freedom of the force sensor.}
\label{tab:freq_linewidth}
\begin{tabular}{ccc}
\hline\hline\addlinespace
Mode & Frequency (Hz) & Linewidth (Hz) \\
\midrule
$x$-translation & 4.0  & 0.48 \\
$y$-translation & 12.6 & 0.37 \\
$z$-translation & 21.2 & 0.42 \\
$x$-rotation    & 68.0 & 0.44 \\
$y$-rotation    & 47.4 & 0.25 \\
$z$-rotation    & 10.1 & 0.44 \\
\hline\hline
\end{tabular}
\end{table}
\begin{figure*}[htbp]
    \centering
    \hypertarget{exp-sys5}{}
    \includegraphics[width=\textwidth,]{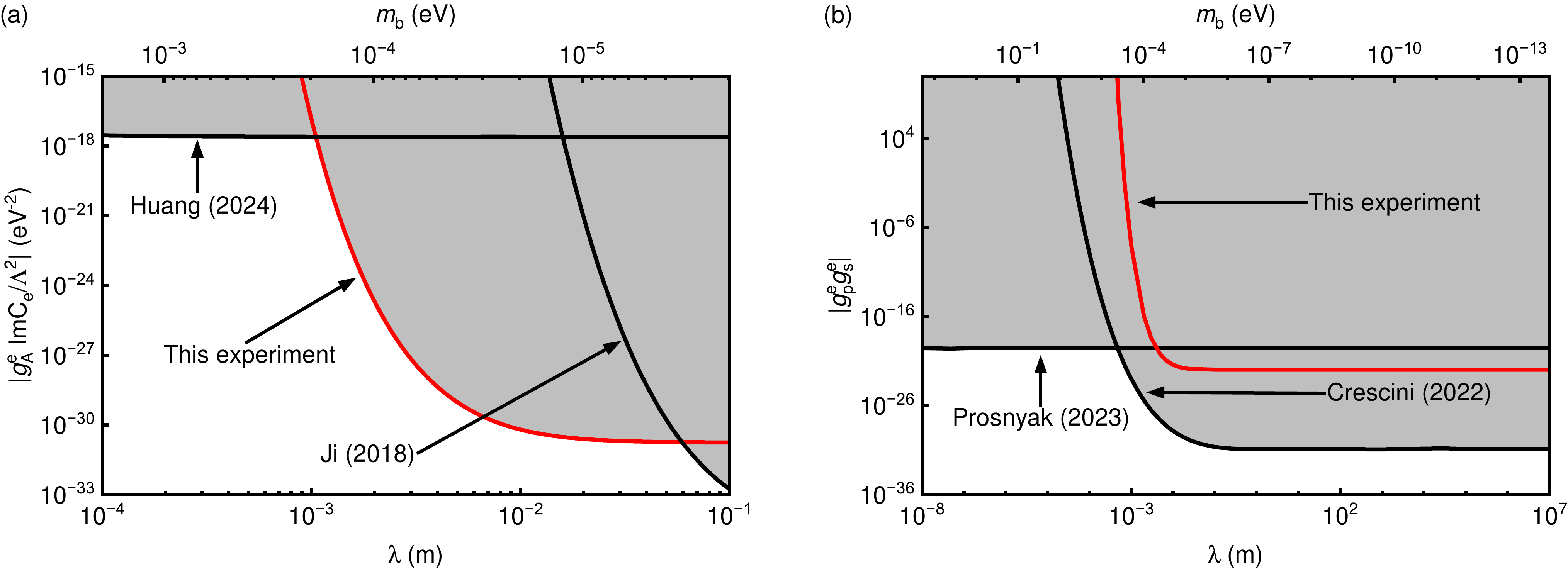}
	\caption{(Color online). Constraints on fundamental coupling constants obtained from experiments. The black lines represent the upper limits established by previous experiments, while the red lines show the upper bounds obtained from our experiment. (a) Constraints on the product $|g_\text{A}^e \mathrm{Im}(C_e)/\Lambda^2|$ as a function of the interaction range $\lambda$ (or the boson mass $m_\text{b}$). At $\lambda = 1.6 \times 10^{-2}~\rm{m}$, the upper limit of the coupling constant is $|g_{\rm A}^{\rm e} \mathrm{Im}(C_e) / \Lambda^{2}| \leq 3.17 \times 10^{-31}~\rm{eV}^{-2}$, which improves the previous result by up to 12 orders of magnitude.  (b) Constraints on the product $|g_\text{P}^eg_\text{S}^e|$ as a function of $\lambda$ (or $m_\text{b}$). }
\end{figure*}
\textit{The
details of the system’s vibration modes}\textbf{---}We measured the frequencies of the sensor's six degrees of freedom, specifically the translations along and rotations about the $x$, $y$, and $z$ axes. The measured frequencies and their corresponding linewidths are summarized in Table~\hyperref[tab:freq_linewidth]{II}. To identify the eigenmodes, we used a frequency-sweep method: by applying electromagnetic signals over a continuous frequency range, we excited the sensor and observed its maximum response at the eigenfrequencies. The corresponding motion directions were then characterized using an optical microscope. For example, to identify the $z$-axis translational mode, we tracked the motion trajectories of multiple distinct points on the sensor, confirming that all points moved uniformly along the $z$ axis. The $x$ and $y$ axis translational modes were identified using the same approach. Similarly, the rotational modes were distinguished by analyzing the distinct motion characteristics across the sensor. For instance, during the identification of the rotation about the $x$ axis, microscopic observation revealed that all tracked points followed small arc trajectories around the $x$ axis. In this way, we well distinguished the rotational modes about the $x$, $y$, and $z$ axes from their translational counterparts.

\textit{Fundamental coupling constraints from experimental data} \textbf{---} Our experiment sets constraints on two distinct types of new physics interactions  between electrons in Fig. \hyperlink{exp-sys5}{5}. The first is mediated by a spin-1 boson, parameterized by the combination $|g_{\rm A}^{\rm e} \mathrm{Im}(C_e) / \Lambda^{2}|$, where $g_{\rm A}^{\rm e}$ is the axial-vector coupling, $\mathrm{Im}(C_e)$ is the pseudotensor coupling, and $\Lambda$ is the ultraviolet cutoff scale. The second is mediated by a spin-0 boson, characterized by the product $|g_{\rm P}^{\rm e} g_{\rm S}^{\rm e}|$, where $g_{\rm S}^{\rm e}$ is the scalar coupling and $g_{\rm P}^{\rm e}$ is the pseudoscalar coupling.

\twocolumngrid                    

\end{document}